\documentclass[12pt]{article}
\usepackage{amsmath}
\usepackage{graphicx,psfrag,epsf}
\usepackage{enumerate}
\usepackage{natbib}
\usepackage{url} 
\usepackage{verbatim}
\usepackage{amssymb}
\usepackage{amsfonts}
\usepackage{epsfig}
\usepackage{epstopdf}
\usepackage{subfigure}
\usepackage{color}
\usepackage{bm}

\newcommand{\blind}{0}

\newcommand{\eps}{\epsilon}
\newcommand{\be}{\begin{eqnarray}}
\newcommand{\ee}{\end{eqnarray}}
\newcommand{\ba}{\begin{eqnarray*}}
\newcommand{\ea}{\end{eqnarray*}}
\newcommand {\abs}[1]{| {#1} |}

\newcommand{\sign}{\mbox{sign}}

\begin{document}

\def\spacingset#1{\renewcommand{\baselinestretch}%
{#1}\small\normalsize} \spacingset{1}


\if0\blind
{
  \title{\bf Fully Bayesian Estimation and Variable Selection in Partially Linear Wavelet Models}
  \author{Norbert Rem\'enyi 
    \\
    School of Industrial and Systems Engineering \\
    Georgia Institute of Technology}
  \date{}
  \maketitle
} \fi

\if1\blind
{
  \bigskip
  \bigskip
  \bigskip
  \begin{center}
    {\LARGE\bf Title}
\end{center}
  \medskip
} \fi

\bigskip
\begin{abstract}
In this paper we propose a wavelet-based methodology for estimation and variable selection in partially linear models. The inference is conducted in the wavelet domain, which provides a sparse and localized decomposition appropriate for nonparametric components with various degrees of smoothness. A hierarchical Bayes model is formulated on the parameters of this representation, where the estimation and variable selection is performed by a Gibbs sampling procedure. For both the parametric and nonparametric part of the model we are using point-mass-at-zero contamination priors with a double exponential spread distribution. Only a few papers in the area of partially linear wavelet models exist, and we show that the proposed methodology is often superior to the existing methods with respect to the task of estimating model parameters. Moreover, the method is able to perform Bayesian variable selection by a stochastic search for the parametric part of the model.
\end{abstract}

\noindent%
\vfill

\newpage
\spacingset{1.45} 
\section{Introduction} \label{sec_intro_gswapalim}
In the present paper we consider a novel Bayesian approach for solving the following regression problem
\be
Y_i = \bm{x}_i^T \bm{\beta} + \bm{f}(t_i) + \varepsilon_i, \quad i=1,\dots, n,
\label{plm_problem}
\ee
where $t_i$, $i=1, \ldots,n$, are equispaced sampling points, $\bm{x}_i$, $i=1, \ldots,n$, are known $p$-dimensional design points, $\bm{\beta}$ is an unknown $p$-dimensional parameter vector, $\bm{f}$ is an unknown and potentially non-smooth function, and the random errors $\varepsilon_i$ are i.i.d. normal, with zero mean and variance $\sigma^2$. The model can be written in matrix-vector form as
\be
\bm Y = \bm X \bm{\beta} + \bm f + \bm{\varepsilon}.
\label{plm_problemv}
\ee
Our interest is to simultaneously estimate the unknown parameter vector $\bm{\beta}$ and nonparametric function $\bm{f}$ using the observations $\bm{Y}$. Another task is to identify important (non-zero) components of $\bm{\beta}$, that is to perform dimension reduction via variable selection on $\bm{\beta}$.

The model in (\ref{plm_problem}) is called a partially linear model (PLM) in the literature. \citet{Engle1986} were among the first to use PLM to analyze electricity sales data. The model is semiparametric in nature because it combines parametric (linear) and nonparametric parts. In this paper we consider a model with one nonparametric part in it. The monograph by \citet{Hardle2000} discusses the general PLM model extensively.

Several approaches are proposed in the literature to represent the nonparametric component $\bm f$ of the model in (\ref{plm_problemv}). These all build on existing nonparametric regression techniques, such as the kernel method, the local linear method (local polynomial or trigonometric polynomial techniques), or splines. In the most recent papers, wavelets are used \citep{Chang2004,Fadili2005b,Qu2006,Gannaz2007,Ding2011}, which allows the nonparametric component to be parsimoniously represented by a limited number of coefficients. The wavelet representation can include a wide variety of nonparametric parts, including non-smooth signals, and reduces the bias in estimating the parametric component.

In this paper we consider the latter approach and use the wavelet decomposition to represent $\bm f$. We use the Bayesian approach to formulate a hierarchical model in the wavelet domain and estimate its parameters. Only a few papers used wavelets in the partially linear model context, and besides \citet{Qu2006}, all used a penalized least squares estimation procedure. Therefore, using a fully Bayesian approach can be of interest.

After applying a linear and orthogonal wavelet transform, the model in (\ref{plm_problem}) becomes
\be
d_{jk} = \bm{u}_{jk}^T \bm{\beta} + \theta_{jk}+\tilde{\varepsilon}_{jk},
\label{plm_model0}
\ee
where $d_{jk}$, $\theta_{jk}$ and $\tilde{\varepsilon}_{jk}$ are the wavelet coefficients (at resolution $j$ and location $k$) corresponding to $\bm{Y}$, $\bm{f}$ and $\bm{\varepsilon}$, and $\bm{U} = \bm{W}\bm{X}$, where $\bm{W}$ is an orthogonal matrix implementing the wavelet transform. In a matrix-vector form,
\ba
\bm{W} \bm Y = \bm{W} \bm X \bm{\beta} + \bm{W} \bm f + \bm{W} \bm{\varepsilon},
\ea
which becomes
\be
\bm d = \bm{U} \bm{\beta} + \bm{\theta} + \bm{\tilde{\varepsilon}}.
\label{plm_model0v}
\ee
Note that because of the orthogonality of $\bm{W}$, $\bm{\tilde{\varepsilon}} \sim {\cal N} (0, \sigma^2I)$. Due to the whitening property of wavelet transforms \citep{Flandrin1992}, we can assume independence of the coefficients $d_{jk}$. To estimate $\beta_i$ and $\theta_{jk}$ in model (\ref{plm_model0}) in a Bayesian fashion, we build on results from the Bayesian linear models and wavelet regression literature.

To estimate $\theta_{jk}$ in a simple nonparametric regression model $Y_i = \bm{f}(t_i) + \varepsilon_i$, Bayesian shrinkage rules have been proposed in the literature by many authors. By a shrinkage rule, we mean that the observed wavelet coefficients $d$ are replaced with their shrunken version $\hat{\theta}=\delta(d)$. The traditional Bayesian models consider a prior distribution on a generic wavelet coefficient $\theta$ as
\be
\pi(\theta)=\eps\delta_0+(1-\eps)\xi(\theta) \label{plm_mixture},
\ee
where $\delta_0$ is a point mass at zero, $\xi$ is a symmetric about 0 and unimodal distribution, and $\eps$
is a fixed parameter in [0,1], usually level dependent, that controls the amount of shrinkage for values of $d$ close to 0. This type of model was considered by \citet{Abramovich1998}, \citet{Vidakovic1998a}, \citet{Vidakovic2001}, and \citet{Johnstone2005a}, among others. 

The mixture prior approach was also utilized in Bayesian estimation and variable selection of linear models, $\bm Y = \bm X \bm{\beta} + \bm{\varepsilon}$, where a mixture prior is specified on parameters $\beta_i$. This type of model was considered for example by \citet{George1993,George1997}, and \citet{Yuan2004,Yuan2005}. It is natural to combine these approaches; therefore, we build on these modeling ideas to formulate a fully Bayesian model in the partially linear model context.

The paper is organized as follows. Section \ref{sec_model_gswapalim} formalizes the Bayesian model and presents some results related to it. In Section \ref{sec_gibbs_gswapalim} we explain the estimation through a Gibbs sampling procedure developed for the hierarchical model. Section \ref{sec_sim_gswapalim} discusses the selection of hyperparameters, contains simulations and comparisons to existing methods, and discusses how variable selection can be performed. Conclusions and discussion are provided in Section \ref{sec_conclusions_gswapalim}.

\section{Hierarchical Model} \label{sec_model_gswapalim}
In this section we propose a hierarchical model in which we use a mixture prior approach for both the parametric and the nonparametric components of the partially linear model. 
Let us consider the following hierarchical Bayesian model for a partially linear model in the wavelet domain (\ref{plm_model0v}):
\be
\bm d|\bm\beta,\bm\gamma,\bm\theta, \sigma^2 &\sim& {\cal N} (\bm{U}_{\bm{\gamma}}\bm\beta_{\bm{\gamma}}+\bm\theta, \sigma^2I) \nonumber \\
\sigma^2                        &\sim& {\cal IG} (a_1,b_1) \nonumber \\
\beta_i|\gamma_i,\tau_{\beta}   &\sim& (1-\gamma_i)\delta_0+\gamma_i{\cal DE}(\tau_{\beta}),\quad i=1,\ldots,p
\nonumber \\
\theta_{jk}|z_{jk},\tau_{\theta}   &\sim& (1-z_{jk})\delta_0+z_{jk}{\cal DE}(\tau_{\theta}) \nonumber \\
\gamma_i|q                &\sim& {\cal B}er (q),\quad i=1,\ldots,p \nonumber \\
z_{jk}|\eps_j             &\sim& {\cal B}er (\eps_j) \nonumber \\
q                         &\sim& {\cal U} (0,1) \nonumber \\
\eps_j                    &\sim& {\cal U} (0,1),
\label{plm_model}
\ee
where $j$ pertains to the resolution level of $d_{jk}$ and ${\cal N}$, ${\cal IG}$, ${\cal DE}$, ${\cal B}er$, and ${\cal U}$ stand for the normal, inverse gamma, double exponential, Bernoulli, and uniform distributions, respectively. Index $i$ refers to the regression coefficients in $\bm{\beta}$. Note that $\bm\gamma$ is an indicator vector of binary elements; therefore, subscript $\bm\gamma$ indicates that only those columns or elements of $\bm{U}$ and $\bm\beta$ with the corresponding $\bm\gamma$ element of 1 are included.

Note that the model in (\ref{plm_model}) uses the well-established mixture prior on $\theta_{jk}$ with a point mass at zero, which accounts for the sparsity of the nonparametric part in the wavelet domain. Wavelet coefficients with large magnitudes are captured by the spread part of the mixture prior, for which we propose the double exponential or Laplace distribution with variance $2/\tau_{\theta}^2$. The double exponential distribution is a popular choice for the spread part. It models wavelet coefficients with large energies and was used by several authors, for example \citet{Vidakovic2001}, and \citet{Johnstone2005a}. The mixture prior on $\theta_{jk}$ is specified levelwise, for each dyadic level $j$; however, the scale parameter $\tau_{\theta}$ is global. This serves the purpose of parsimony and contributes to the ease of estimation. Here $z_{jk}$ is a latent variable indicating whether our parameter $\theta_{jk}$ is coming from a point mass at zero ($z_{jk}=0$) or from a double exponential part ($z_{jk}=1$), with prior probability of $1-\eps_j$ or $\eps_j$, respectively. For the prior probability $\eps_j$ we assume a ``noninformative'' uniform prior. The uniform ${\cal U}$(0,1) prior is equivalent to a beta ${\cal B}e(1,1)$ distribution, which is a conjugate prior for the Bernoulli distribution. 

In our model we naturally propose the same mixture prior to model the regression parameters $\beta_i,~ i=1,\ldots,p$. \citet{Yuan2004,Yuan2005} used this prior in the Bayesian variable selection context for linear models. In case $\gamma_i=0$ the model forces $\beta_i=0$ and if $\gamma_i=1$ then $\beta_i$ is modeled with a double exponential prior accommodating large regression coefficients. For the elements of binary vector $\bm\gamma$ we use the Bernoulli prior with common parameter $q$. This prior assumes that each predictor enters the model independently with prior probability $q$. Although it does not take into account the possible correlation between the predictors, this type of prior works well in practice, and it was used by \citet{George1993} and \citet{George2000}, to name a few. Unlike \citet{George1993}, who prespecified $q$, we introduce another level of hierarchy by assuming a uniform ``noninformative'' prior on $q$. Since it is not clear, in general, how to specify $q$, it makes sense to put a prior distribution on the parameter, instead of using $q=1/2$, which is a common suggestion in practice. As opposed to the fully Bayesian approach, \citet{George2000} used the empirical Bayes approach to estimate $q$.

Parameter $\sigma^2$ represents the common noise variance for each resolution level on which we specified a conjugate inverse gamma prior. Spread parameters $\tau_{\theta}$ and $\tau_{\beta}$ will be given priors after a reformulated version of the model (\ref{plm_model}) is discussed.

The hierarchical model in (\ref{plm_model}) is not conjugate; however, with additional transformations, derivations and computational techniques, it is possible to develop a fast Gibbs sampling algorithm for updating of its parameters. Note that a standard approach for handling the double exponential prior in Markov chain Monte Carlo (MCMC) computations of hierarchical models is to represent the double exponential distribution as a scale mixture of normal distributions \citep{Andrews1974}. This approach is used for example in Bayesian LASSO variable selection, where the double exponential prior (without point mass) is used on the regression parameters \citep{Park2008}. Here we will only use the scale mixture approach for the double exponential prior on $\beta_i$. This introduces an additional parameter $v_i$ corresponding to each $\beta_i$, which needs to be updated. Using the scale mixture representation, the model in ($\ref{plm_model}$) becomes
\be
\bm d|\bm\beta,\bm\gamma,\bm\theta, \sigma^2 &\sim& {\cal N} (\bm{U}_{\bm{\gamma}}\bm\beta_{\bm{\gamma}}+\bm\theta, \sigma^2I) \nonumber \\
\sigma^2                    &\sim& {\cal IG} (a_1,b_1) \nonumber \\
\beta_i|\gamma_i,v_i,\eta^2 &\sim& (1-\gamma_i)\delta_0+\gamma_i\,{\cal N}(0,v_i\eta^2),\quad i=1,\ldots,p
\nonumber \\
v_i                         &\sim& {\cal E}xp(1),\quad i=1,\ldots,p \nonumber \\
\theta_{jk}|z_{jk},\tau_{\theta}     &\sim& (1-z_{jk})\delta_0+z_{jk}{\cal DE}(\tau_{\theta}) \nonumber \\
\gamma_i|q                  &\sim& {\cal B}er (q),\quad i=1,\ldots,p \nonumber \\
z_{jk}|\eps_j               &\sim& {\cal B}er (\eps_j) \nonumber \\
q                           &\sim& {\cal U} (0,1) \nonumber \\
\eps_j                      &\sim& {\cal U} (0,1) \nonumber \\
\eta^2                      &\sim& {\cal IG} (a_2,b_2) \nonumber \\
\tau_{\theta}               &\sim& {\cal G}a (a_3,b_3)
\label{plm_model2}
\ee

In the model above $\eta=\sqrt{2}/\tau_{\beta}$. If we integrate out $v_i$s from ($\ref{plm_model2}$), we get back the model in ($\ref{plm_model}$), which follows from the scale mixture representation of the double exponential distribution. For the spread parameters $\eta^2$ and $\tau_{\theta}$, inverse gamma and gamma priors are specified in the model, which turn out to be conjugate.

For parameters $\theta_{jk}$ it is possible to derive the full conditional distributions without resorting to the scale mixture representation. This improves the speed of the Gibbs sampling algorithm. In order to do this, we first discuss some results related to model ($\ref{plm_model2}$), which are instrumental in developing the Gibbs sampler.

First let $d^{\star}_{jk}=d_{jk}-(\bm{U}_{\bm{\gamma}}\bm\beta_{\bm{\gamma}})_{jk}$ from which it follows that $d^{\star}_{jk} \sim {\cal N} (\theta_{jk}, \sigma^2)$. In the following notation $d^{\star}$ refers to an arbitrary $d^{\star}_{jk}$ and the mean $\theta$ stands for the corresponding $\theta_{jk}$. If we consider a ${\cal N} (\theta, \sigma^2)$ likelihood $f(d^{\star}|\theta,\sigma^2)$ and elicit a double exponential ${\cal DE} (\tau)$ prior $p_1(\theta|\tau)$ on the $\theta$, the marginal distribution becomes
\be
m(d^{\star}|\sigma^2,\tau)  &=& \frac{\tau}{2}e^{\frac{\sigma^2\tau^2}{2}}\left\{
                e^{-d^{\star}\tau}\Phi\left(\frac{d^{\star}}{\sigma}-\tau\sigma\right)+
                e^{d^{\star}\tau}\Phi\left(-\frac{d^{\star}}{\sigma}-\tau\sigma\right) \right\} \label{plm_marg},
\ee
and the posterior distribution of $\theta$ becomes
\be
&& h(\theta|d^{\star},\sigma^2,\tau)= \nonumber \\
&& = \begin{cases}
\displaystyle \frac{e^{-d^{\star}\tau}}{e^{-d^{\star}\tau}\Phi\left(\frac{d^{\star}}{\sigma}-\tau\sigma\right)+
e^{d^{\star}\tau}\Phi(-d^{\star}/\sigma-\tau\sigma)}
\frac{1}{\sigma}\phi\left(\frac{\theta-(d^{\star}-\sigma^2\tau)}{\sigma}\right), & \theta \geq 0 \nonumber \\
\displaystyle \frac{e^{d^{\star}\tau}}{e^{-d^{\star}\tau}\Phi\left(\frac{d^{\star}}{\sigma}-\tau\sigma\right)+
e^{d^{\star}\tau}\Phi\left(-\frac{d^{\star}}{\sigma}-\tau\sigma\right)}
\frac{1}{\sigma}\phi\left(\frac{\theta-(d^{\star}+\sigma^2\tau)}{\sigma}\right),
& \theta < 0 \label{plm_post}
\end{cases}, \\
\ee
where $\phi$ and $\Phi$ respectively denote the pdf and cdf of the standard normal distribution. For derivations of these results, see Appendix. From the representation in (\ref{plm_post}) we can see that the posterior distribution is a mixture of truncated normals, which will be utilized in the Gibbs sampling algorithm. If we consider the mixture prior $p(\theta|\tau) = (1-\eps_j)\delta_0 + \eps_j p_1(\theta|\tau)$ on $\theta$ in (\ref{plm_model}), we obtain the posterior distribution as
\be
\pi(\theta|d^{\star},\sigma^2,\tau)  &=& \frac{f(d^{\star}|\theta,\sigma^2)p(\theta|\tau)}
    {\int_{-\infty}^{\infty}{f(d^{\star}|\theta,\sigma^2)p(\theta|\tau)d\theta}} \nonumber \\
                                &=& \frac{(1-\eps_j)f(d^{\star}|\theta,\sigma^2)\delta_0+
    \eps_j f(d^{\star}|\theta,\sigma^2)p_1(\theta|\tau)}{(1-\eps_j)f(d^{\star}|0,\sigma^2)+\eps_j m(d^{\star}|\sigma^2,\tau)} \nonumber \\
                                &=& \frac{(1-\eps_j)f(d^{\star}|0,\sigma^2)\delta_0+
    \eps_j m(d^{\star}|\sigma^2,\tau)h(\theta|d^{\star},\sigma^2,\tau)}{(1-\eps_j)f(d^{\star}|0,\sigma^2)+
    \eps_j m(d^{\star}|\sigma^2,\tau)} \nonumber \\
                                &=& (1-p_j)\delta_0+p_j h(\theta|d^{\star},\sigma^2,\tau), \label{plm_post2}
\ee
where $f(d^{\star}|0,\sigma^2)$ is the normal distribution with mean $\theta=0$ and variance $\sigma^2$, and
\be
p_j = \frac{\eps_j m(d^{\star}|\sigma^2,\tau)}{(1-\eps_j)f(d^{\star}|0,\sigma^2)+\eps_j m(d^{\star}|\sigma^2,\tau)} \label{plm_p} \\
\nonumber
\ee
is the mixing weight. Thus, the posterior distribution of $\theta$ is a mixture of point mass at zero and a mixture of truncated normal distributions $h(\theta|d^{\star},\sigma^2,\tau)$ with mixing weight $p_j$.

\section{Gibbs sampling scheme} \label{sec_gibbs_gswapalim}
To conduct posterior inference on the parameters $\theta_{jk}$ and $\beta_i$, we adopt a standard Gibbs sampling procedure. Gibbs sampling is an iterative algorithm that simulates from a joint posterior distribution through iterative simulation of the full conditional distributions. For more details on Gibbs sampling see \citet{Casella1992} or \citet{Robert1999}. For the model in (\ref{plm_model2}), full conditionals for all parameters can be determined exactly. We build on results given as (\ref{plm_post}), (\ref{plm_post2}) and results derived by \citet{Yuan2004}. Derivations of the results in this section are deferred to Appendix.

Next we will find full conditional distributions and updating schemes for parameters $\gamma_i$, $\beta_i$, $v_i$, $\eta^2$, $q$, $\sigma^2$, $z_{jk}$, $\eps_j$, $\theta_{jk}$, and $\tau_{\theta}$, which are necessary to run the Gibbs sampler. Specification of the hyperparameters $a_1$, $b_1$, $a_2$, $b_2$, $a_3$ and $b_3$ will be done in Section \ref{section_hyper_gswapalim}.

\subsection{Updating $\gamma_i$, $\beta_i$ and $v_i$}
In each Gibbs sampling iteration we first update the block $(\gamma_i,\beta_i)$ by updating $\gamma_i$ and $\beta_i$ for $i=1,\ldots,p$, and then we generate $v_i$ for $i=1,\ldots,p$.

\subsubsection{Updating $\gamma_i$ and $\beta_i$ as a block}
Here we follow the results of \citet{Yuan2004} and we get
\ba
P(\gamma_i=1|\bm{d},\bm{\theta},\sigma^2,\eta^2,\bm{\beta}^{[-i]},v_i,\bm{\gamma}^{[-i]})=\frac{1}{1+\frac{f(\bm{d}|
\bm{\theta},\sigma^2,\eta^2,
\bm{\beta}^{[-i]},v,\bm{\gamma}^{[-i]},\gamma_i=0)P(\bm{\gamma}^{[-i]},\gamma_i=0)}
{f(\bm{d}|\bm{\theta},\sigma^2,\eta^2,\bm{\beta}^{[-i]},v,\bm{\gamma}^{[-i]},\gamma_i=1)
P(\bm{\gamma}^{[-i]},\gamma_i=1)}},
\ea
where
\ba
f(\bm{d}|\bm{\theta},\sigma^2,\eta^2,\bm{\beta}^{[-i]},v,\bm{\gamma}^{[-i]},\gamma_i=0)=
\left(\frac{1}{\sqrt{2\pi\sigma^2}}\right)^n \exp\left\{-\frac{\bm{Z}'\bm{Z}}{2\sigma^2}\right\},
\ea
and
\ba
&& f(\bm{d}|\bm{\theta},\sigma^2,\eta^2,\bm{\beta}^{[-i]},v,\bm{\gamma}^{[-i]},\gamma_i=1)= \\
&& \left(\frac{1}{\sqrt{2\pi\sigma^2}}\right)^n
\exp\left\{-\frac{\bm{Z}'\bm{Z}}{2\sigma^2}\right\}\sqrt{\frac{\sigma^2}{v_i\eta^2 \bm{U}'_i \bm{U}_i+\sigma^2}}
\exp\left\{\frac{v_i\eta^2(\bm{Z}'\bm{U}_i)^2}{2\sigma^2(v_i\eta^2 \bm{U}'_i \bm{U}_i+\sigma^2)} \right\}.
\ea
Note that
\ba
\bm{Z} = \bm{d}-\bm{U}_{\bm{\gamma}^{[-i]},\gamma_i=0} \bm{\beta}_{\bm{\gamma}^{[-i]},\gamma_i=0}-\bm{\theta},
\ea
and
\ba
\frac{P(\bm{\gamma}^{[-i]},\gamma_i=0)}{P(\bm{\gamma}^{[-i]},\gamma_i=1)}=\frac{1-q^{(l-1)}}{q^{(l-1)}}.
\ea
Here the notation $\bm{\gamma}^{[-i]}$ and $\bm{\beta}^{[-i]}$ refers to vectors $\bm{\gamma}$ and $\bm{\beta}$ without the $i^{th}$ element and $\bm{U}_i$ indicates the $i^{th}$ column of matrix $\bm{U}$. Therefore, in the $l^{th}$ iteration of the Gibbs sampling, update $\gamma_i$ as a Bernoulli random variable with probabilities given
\be
\gamma^{(l)}_i =
\begin{cases}
0, & \mbox{wp.} \quad \displaystyle
    1-P\left(\gamma_i=1\big|\bm{d},\bm{\theta}^{(l-1)},{\sigma^2}^{(l-1)},{\eta^2}^{(l-1)},{\bm{\beta}^{[-i]}}^{(l)},
    v_i^{(l-1)},{\bm{\gamma}^{[-i]}}^{(l)}
    \right) \\
1, & \mbox{wp.} \quad \displaystyle
    P\left(\gamma_i=1\big|\bm{d},\bm{\theta}^{(l-1)},{\sigma^2}^{(l-1)},{\eta^2}^{(l-1)},{\bm{\beta}^{[-i]}}^{(l)},
    v_i^{(l-1)},{\bm{\gamma}^{[-i]}}^{(l)}
    \right) \nonumber
\end{cases}. \\
\ee

\noindent Then it is straightforward to update $\beta_i$ as
\be
\beta^{(l)}_i \sim
\begin{cases}
\delta_0(\beta_{i}), & \mbox{if} \quad \gamma^{(l)}_i=0 \\
{\cal N} \left(\frac{v_i^{(l-1)}{\eta^2}^{(l-1)}(\bm{Z}'\bm{U}_i)^2}{v_i^{(l-1)}{\eta^2}^{(l-1)} \bm{U}'_i \bm{U}_i+
{\sigma^2}^{(l-1)}}, \frac{v_i^{(l-1)} {\eta^2}^{(l-1)} {\sigma^2}^{(l-1)}}{v_i^{(l-1)} {\eta^2}^{(l-1)}
\bm{U}'_i \bm{U}_i+{\sigma^2}^{(l-1)} } \right), & \mbox{if} \quad \gamma^{(l)}_i=1 \nonumber
\end{cases}. \\
\ee
Note that in the above equation $\bm{Z} = \bm{d}-\bm{U}_{\bm{\gamma}^{[-i]},\gamma_i=0}\bm{\beta}_{\bm{\gamma}^{[-i]},\gamma_i=0}-\bm{\theta}$ in which we substitute ${\bm{\gamma}^{[-i]}}^{(l)}$, $\bm{\beta}^{(l)}$ and $\bm{\theta}^{(l-1)}$. Also, $\delta_0(\beta_i)$ is a point mass distribution at zero, which is equivalent to $\beta_i=0$.

\subsubsection{Updating $v_i$}
For the scale mixture of normals representation of the double exponential distribution, we placed an exponential prior on $v_i$ in model (\ref{plm_model2}). We update $v_i$ depending on the value of the latent variable $\gamma_i$, whether $\beta_i$ comes from a point mass or a normal prior. The updating scheme for $v_i$ is
\be
v^{(l)}_i \sim
\begin{cases}
{\cal E}xp(1), & \mbox{if} \quad \gamma^{(l)}_i=0 \\
{\cal GIG} \left(2,{\beta^2_i}^{(l)}/{\eta^2}^{(l-1)},1/2 \right), & \mbox{if} \quad \gamma^{(l)}_i=1
\end{cases},
\ee
where ${\cal GIG}(a,b,p)$ denotes the generalized inverse Gaussian distribution \citep[p.284]{Johnson1994} with probability density function
\ba
f(x|a,b,p)=\frac{(a/b)^{p/2}}{2K_p(\sqrt{ab})}x^{p-1}e^{-(ax+b/x)/2}, \quad x>0;a,b>0.
\ea
Here $K_p$ denotes the modified Bessel function of the third kind. Simulation of ${\cal GIG}$ random variates is available through a MATLAB$^\copyright$ implementation ``randraw'' based on \citet{Dagpunar1989}.

\subsection{Updating $\eta^2$, $q$, $\eps_{j}$ and $\sigma^2$}
Using a conjugate ${\cal IG} (a_2,b_2)$ prior on $\eta^2$ results in an inverse gamma full conditional distribution. Therefore, update $\eta^2$ as
\be
{\eta^2}^{(l)} \sim {\cal IG} \left( a_2+1/2 \sum_i {\gamma_i}^{(l)},
\left[1/b_2+1/2 \sum_i{\left( {\gamma_i}^{(l)} {\beta^2_i}^{(l)} / v_i^{(l)} \right)} \right]^{-1} \right).
\ee

Parameter $q$ has a conjugate ${\cal B}e(1,1)$ prior. This results in a full conditional distributed as beta,
\be
q^{(l)} \sim {\cal B}e \left(1+\sum_i {\gamma_i}^{(l)}, 1+\sum_i \left(1-{\gamma_i}^{(l)}\right) \right).
\ee

Similarly, parameter $\eps_j$ is given a conjugate ${\cal B}e(1,1)$ prior, and the update is
\be
\eps^{(l)}_j \sim {\cal B}e \left(1+\sum_k z^{(l)}_{jk}, 1+\sum_k \left(1-z^{(l)}_{jk}\right) \right).
\ee
Note that other choices from the ${\cal B}e(\alpha,\beta)$ family are possible for the prior of $\eps_j$ and $q$, similarly. However, we used the noninformative choice $\alpha=1$ and $\beta=1$ to facilitate data-driven estimation of $\eps_j$ and $q$.

Using a conjugate ${\cal IG} (a_1,b_1)$ prior on $\sigma^2$ also results in an inverse gamma full conditional distribution. This leads to an update for $\sigma^2$ as
\be
{\sigma^2}^{(l)} \sim {\cal IG} \left( a_1+n/2, \left[1/b_1 + \bm{Z}'\bm{Z}/2 \right]^{-1} \right),
\ee
where $\bm{Z} = \bm{d}-\bm{U}_{\bm{\gamma}^{(l)}}\bm{\beta}^{(l)}_{\bm{\gamma}^{(l)}}-\bm{\theta}^{(l-1)}$ and $n=2^J-2^{J_0}$ denotes the sample size. $J-1$ and $J_0$ refer to the finest and coarsest levels in the wavelet decomposition, respectively.

\subsection{Updating $z_{jk}$}
We saw in model (\ref{plm_model2}) that latent variable $z_{jk}$ has a Bernoulli prior with parameter $\eps_j$. Its full conditional distribution remains Bernoulli with parameter $p_j$ as in (\ref{plm_p}). Thus, the latent variable $z_{jk}$ is updated as follows:
\be
z^{(l)}_{jk}=
\begin{cases}
0, & \mbox{wp.} \quad \displaystyle
    \frac{\left(1-\eps^{(l-1)}_j\right)f\left(d^{\star}_{jk}\big|0,{\sigma^2}^{(l)}\right)}
    {\left(1-\eps^{(l-1)}_j\right)f\left(d^{\star}_{jk}\big|0,{\sigma^2}^{(l)}\right)+
    \eps^{(l-1)}_j m\left(d^{\star}_{jk}\big|{\sigma^2}^{(l)},\tau_{\theta}^{(l-1)} \right)} \nonumber \\
1, & \mbox{wp.} \quad \displaystyle
    \frac{\eps^{(l-1)}_j m\left(d^{\star}_{jk}\big|{\sigma^2}^{(l)},\tau_{\theta}^{(l-1)} \right)}
    {\left(1-\eps^{(l-1)}_j\right)f\left(d^{\star}_{jk}\big|0,{\sigma^2}^{(l)}\right)+
    \eps^{(l-1)}_j m\left(d^{\star}_{jk}\big|{\sigma^2}^{(l)},\tau_{\theta}^{(l-1)} \right)}
\end{cases} \\
\ee
where $d^{\star}_{jk}=d_{jk}-\left( \bm{U}_{\bm{\gamma}^{(l)}}\bm{\beta}^{(l)}_{\bm{\gamma}^{(l)}} \right)_{jk}$.

\subsection{Updating $\theta_{jk} $}
We approach updating $\theta_{jk}$ in a novel way. As we mentioned before, the common approach for handling the double exponential prior in hierarchical models is the scale mixture representation. This approach, however, introduces an additional parameter corresponding to each $\theta_{jk}$, which needs to be updated. This adds $2^J-2^{J_0}$ new parameters. A faster and more direct method to update $\theta_{jk}$ is possible by using results in (\ref{plm_post}) and (\ref{plm_post2}). From the definition of latent variable $z_{jk}$ we can easily see that $\theta_{jk}=0$ if $z_{jk}=0$, because for such $z_{jk}$, $\theta_{jk}$ is distributed as point mass at zero. In case $z_{jk}=1$, $\theta_{jk}$ follows a mixture of truncated normal distributions a posteriori. Therefore, the update for $\theta_{jk}$ is as follows:
\be
\theta^{(l)}_{jk} \sim
\begin{cases}
\delta_0(\theta_{jk}), & \mbox{if} \quad z^{(l)}_{jk}=0 \\
h\left(\theta_{jk}\big|d^{\star}_{jk},{\sigma^2}^{(l)},\tau_{\theta}^{(l-1)}\right), & \mbox{if} \quad z^{(l)}_{jk}=1
\end{cases},
\ee
where $d^{\star}_{jk}=d_{jk}-\left( \bm{U}_{\bm{\gamma}^{(l)}}\bm{\beta}^{(l)}_{\bm{\gamma}^{(l)}} \right)_{jk}$, $\delta_0(\theta)$ is a point mass distribution at zero, and $h(\theta|d^{\star},\sigma^2,\tau_{\theta})$ is a mixture of truncated normal distributions with the density provided in (\ref{plm_post}). Simulating random variables from $h(\theta|d^{\star},\sigma^2,\tau_{\theta})$ is nonstandard, and regular built-in methods fail, because we need to simulate random variables from tails of normal distributions having extremely low probability. The implementation of the updating algorithm is based on vectorizing a fast algorithm proposed by \citet{Robert1995}.

\subsection{Updating $\tau_{\theta}$}
The Gibbs updating scheme is completed with the discussion of how to update $\tau_{\theta}$. In the hierarchical model (\ref{plm_model2}), we impose a gamma prior on the scale parameter of the double exponential distribution. This turns out to be a conjugate problem; therefore, we update $\tau_{\theta}$ by
\be
\tau_{\theta}^{(l)} \sim {\cal G}a \left(a_3+\sum_{j,k}z^{(l)}_{jk},\left[1/b_3+
\sum_{j,k}\left(z^{(l)}_{jk}|\theta^{(l)}_{jk}|\right)\right]^{-1} \right).
\ee
Note that the gamma distribution above is parameterized by its scale parameter.

Now the derivation of the updating algorithm is complete. Implementation of the described Gibbs sampler requires simulation routines for standard distributions such as the gamma, inverse gamma, Bernoulli, beta, exponential, normal, and also specialized routines to simulate from truncated normal, and generalized inverse Gaussian. The procedure was implemented in MATLAB and available from the author.

The Gibbs sampling procedure can be summarized as
\begin{enumerate}[(i)]
\item Choose initial values for parameters
\item Repeat steps (iii) - (xi) for $l=1,\ldots,M$
\item Update the block $(\gamma_i,\beta_i)$ for $i=1,\ldots,p$
\item Update $v_i$ for $i=1,\ldots,p$
\item Update $\eta^2$
\item Update $q$
\item Update $\sigma^2$
\item Update $z_{jk}$ for $j=J_0,\ldots,\log_2(n)-1,~k=0,\ldots,2^j-1$
\item Update $\eps_j$ for $j=J_0,\ldots,\log_2(n)-1$
\item Update $\theta_{jk}$ for $j=J_0,\ldots,\log_2(n)-1,~k=0,\ldots,2^j-1$
\item Update $\tau_{\theta}$.
\end{enumerate}
Note that the updating steps of vectors $\bm v, \bm z, \bm \eps$, and $\bm \theta$ are vectorized in the implementation, which considerably speeds up the computation.

\section{Simulations} \label{sec_sim_gswapalim}
In this section, we apply the proposed Gibbs sampling algorithm and simulate posterior realizations for the model in (\ref{plm_model2}). We will name our method \textit{GS-WaPaLiM}, which is an acronym for Gibbs Sampling Wavelet-based Partially Linear Model (\textit{GS-WaPaLiM}) method. Within each simulation step 20,000 Gibbs sampling iterations were performed, of which the first 5,000 were used for burn-in. We used the sample averages $\hat{\theta}_{jk}=\sum_{l} \theta^{(l)}_{jk}/L$ and $\hat{\beta}_i=\sum_{l} \beta^{(l)}_i/L$ as the usual estimator for the posterior mean. In our set-up, $L=15,000$.

In what follows, we first discuss the selection of the hyperparameters, then compare the estimation performance with other methods on two simulated examples. Finally, variable selection will be demonstrated on an example.

\subsection{Selection of Hyperparameters} \label{section_hyper_gswapalim}
In any Bayesian modeling task, the selection of hyperparameters is critical for good performance of the model.
It is also desirable to have a default choice of the hyperparameters which makes the procedure automatic.

In order to apply the \textit{GS-WaPaLiM} method, we only need to specify hyperparameters $a_1$, $b_1$, $a_2$, $b_2$, $a_3$, and $b_3$ in the hyperprior distributions. The advantage of the fully Bayesian approach is that once the hyperpriors are set, the estimation of parameters $\gamma_i$, $\beta_i$, $v_i$, $\eta^2$, $q$, $\sigma^2$, $z_{jk}$, $\eps_j$, $\theta_{jk}$, and $\tau_{\theta}$ is automatic via the Gibbs sampling algorithm. The selection is governed by the data and hyperprior distributions on the parameters. Another advantage is that the method is relatively robust to the choice of hyperparameters since they influence the model at a higher level of hierarchy.

Critical parameters with respect to the performance of the shrinkage are $\eps_j$ and $q$, which control the strength of shrinkage of $\theta_{jk}$ and $\beta_i$ to zero. In model (\ref{plm_model2}), we placed a uniform prior on these parameters; therefore, the estimation will be governed mostly by the data, which provides a degree of adaptiveness. Parameter $q$ represents the probability that a predictor enters the model a priori. When a priori information is available, it can be incorporated into the model, however, this is rarely the case. In the wavelet regression context, \citet{Abramovich1998} estimated parameter $\eps_j$ by a theoretically justified but somewhat involved method, and in \citet{Vidakovic2001}, the estimation of this parameter depends on another hyperparameter $\gamma$, which is elicited based on empirical evidence. The proposed method provides a better alternative because of its automatic adaptiveness to the underlying nonparametric part of the model.

Another efficient way to elicit the hyperparameters of the model is through the empirical Bayes method performing maximization of the marginal likelihood. This approach was followed by \citet{Qu2006} in the context of estimating partially linear wavelet models. However, the likelihood function is nonconcave; therefore, clever optimization algorithm and carefully set starting values are crucial for the performance of this method. The same method of estimating hyperparameters was used for example by \citet{Clyde1999} and \citet{Johnstone2005a} in the wavelet regression context, and by \citet{George2000} in the linear regression context. Note that for the mixture priors specified on the parametric and nonparametric parts in model (\ref{plm_model2}) the empirical Bayes approach might not be computationally tractable; therefore, the fully Bayesian approach provides a good alternative.

Default specification of hyperparameters $a_1$, $b_1$, $a_2$, $b_2$, $a_3$, and $b_3$ in model (\ref{plm_model2}) is given by the following:
\begin{itemize}
\item We set $a_1=2$, $a_2=2$ and $a_3=1$.
\item Then we compute naive estimators from the data
\ba
\bm{\hat{\beta}}_{OLS}=(\bm{X}'\bm{X})^{-1}\bm{X}'\bm{Y}, \\
\bm{Y}_{\bm{f}}=\bm{Y}-\bm{X}\bm{\hat{\beta}}_{OLS},
\ea
where $\bm{Y}_{\bm{f}}$ is an estimator of the nonparametric part of model (\ref{plm_problemv}), and $\bm{\hat{\beta}}_{OLS}$ is the ordinary least squares estimator for $\bm{\beta}$, although computed from the raw partially linear data.
\item Then we set $b_1=1/\hat{\sigma}^2$, so that the mean of the inverse gamma prior becomes $\hat{\sigma}^2$. We use $\hat{\sigma}^2=MAD/0.6745$, which is the usual robust estimator of the noise variation in the wavelet shrinkage literature \citep{Donoho1994}. Here MAD stands for the median absolute deviation of the wavelet coefficients $d^{\bm{f}}_{jk}$ at the finest level of detail and the constant 0.6745 calibrates the estimator to be comparable with the sample standard deviation. Note that coefficients $d^{\bm{f}}_{jk}$ correspond to $\bm{Y}_{\bm{f}}$, therefore, $d^{\bm{f}}_{jk}=d_{jk}-(\bm{U}\bm{\hat{\beta}}_{OLS})_{jk}$.
\item After this we set $b_3=\hat{\tau}_{\theta}=\left( \sqrt{\max \{(\sigma^2_f-\hat{\sigma}^2),0\}}\right)^{-1}$, which sets the mean of the gamma prior on $\tau_{\theta}$ equal to an estimator of $\tau_{\theta}$. This estimator is adopted from \citet{Vidakovic2001}, where $\sigma^2_f=\text{Var}(\bm{Y}_{\bm{f}})$.
\item Finally we set $b_2=1/\hat{\eta}^2$, so that the mean of the inverse gamma prior is a prespecified value, $\hat{\eta}^2$. Results in the estimation of $\beta_i$s turned out to be somewhat sensitive to $\hat{\eta}^2$ for small sample size and small number of linear predictors. We used $\hat{\eta}^2 = \left(3\max_i\{ |\bm{\hat{\beta}}_{OLS_i}| \} \right)^2$, which specified a prior on $\beta_i$ with large enough variance to work well in practice.
\end{itemize}

\subsection{Simulations and Comparisons with Various Methods}
In this section, we discuss the estimation performance of the proposed \textit{GS-WaPaLiM} method and compare it to three methods from the partially linear wavelet model literature. The first one is the wavelet Backfitting algorithm (\textit{BF}) proposed by \citet{Chang2004}, the second one is the \textit{LEGEND} algorithm proposed by \citet{Gannaz2007} and the last one is the double penalized PLM wavelet estimator (\textit{DPPLM}) by \citet{Ding2011}. A Bayesian wavelet-based algorithm for the same problem was proposed by \citet{Qu2006}. However, we found that the implementation of that algorithm is not robust to different simulated examples and initial values of the empirical Bayes procedure, therefore, we omitted it from our discussion.

The coarsest wavelet decomposition level was $J_0=\lfloor \log_2(\log(n))+1 \rfloor$, as suggested from \citet{Antoniadis2001}. Reconstruction of the theoretical signal was measured by the average mean squared error (AMSE), calculated as
\ba
\textnormal{AMSE} = \frac{1}{Mn}\sum_{m=1}^{M}\sum_{i=1}^{n}\left(\hat{Y}^{(m)}_i-Y_i\right)^2,
\ea
where $M$ is the number of simulation runs, and $Y_i,~i=1,\ldots,n$ are known values of the simulated functions considered. We denote by $\hat{Y}^{(m)}_i,~i=1,\ldots,n$ the estimator from the $m$th simulation run. Note again, that in each of these simulation runs we perform 20,000 Gibbs sampling iterations in order to get the estimators $\hat{\theta}_{jk}$ and $\hat{\beta}_i$. Also note that $\hat{\bm{Y}}=\bm{W}'\bm{\hat{d}}$, where $\bm{\hat{d}}=\bm{U}\bm{\hat{\beta}}+\bm{\hat{\theta}}$. We also assess the performance in estimating the parametric part of the model by AMSE$_{\bm{\beta}}$, calculated as
\ba
\textnormal{AMSE}_{\bm{\beta}} = \frac{1}{M}\sum_{m=1}^{M}\sum_{i=1}^{p}\left(\hat{\beta}^{(m)}_i-\beta_i\right)^2.
\ea

In the following simulation study we also used a modification of the wavelet Backfitting algorithm proposed by \citet{Chang2004}. The original algorithm, denoted as \textit{BF}, uses $\hat{\sigma}\sqrt{2\log(n)}$ as a soft threshold value in each iteration. In the modified algorithm we run the iterative algorithm a second time using the generalized cross-validation threshold as in \citet{Jansen1997}. This simple modification significantly improves the performance of the original algorithm. The method will be denoted as \textit{BFM} in the sequel.

The procedure based on \citet{Gannaz2007}, denoted as \textit{LEGEND}, is a wavelet thresholding based estimation procedure solved by the proposed \textit{LEGEND} algorithm. The formulation of the problem is similar to the one in \citet{Chang2004} and \citet{Fadili2005b}, penalizing only the wavelet coefficients of the nonparametric part, but the solution is faster by recognizing the connection with Huber\textquoteright s M-estimation of a standard linear model with outliers.

The algorithm by \citet{Ding2011} will be denoted as \textit{DPPLM} in the simulations. The authors discuss several simulation results based on how the Lasso penalty parameter $\lambda_2$ was chosen and whether the adaptive Lasso algorithm was used or not in the estimation procedure. It was reported that the \textit{GCV} criteria with adaptive Lasso provided the smallest AMSE results, therefore, that version of the algorithm is used in the present simulations. We will refer to the method as \textit{DPPLM-GCV} in the future.

For comparison purposes we use two simulation examples, one from \citet{Qu2006}, and another one from \citet{Ding2011}. We set the number of replications $M=1000$.

\noindent {\bf Example 1} \\
The first example is based on an example in \citet{Qu2006}. The simulated data are generated from
\ba
Y_i = \bm{x}^T_i\bm{\beta}+\bm{f}(t_i)+\varepsilon_i,~~i=1,\ldots,n,
\ea
where $\varepsilon_i \sim N(0,1)$ and $\bm{\beta}=(0.5,1)'$ with $p=2$. The nonparametric test functions are $\bm{f}(t)=c_j \bm{f}_j(t),~j=1,\ldots,4$, where $\bm{f}_1(t)=$ \tt Blocks\rm, $\bm{f}_2(t)=$ \tt Bumps\rm, $\bm{f}_3(t)=$ \tt Doppler \rm and $\bm{f}_4(t)=$ \tt Heavisine\rm. These are four standard test functions considered by \citet{Donoho1994}. We chose $c_1=3$, $c_2=7$, $c_3=18$ and $c_4=2$ to have reasonable signal-to-noise ratios (SNR).  The test functions were simulated at $n=64,128,256$ and $512$ points, and the nonparametric components were equally spaced in the unit interval. The standard wavelet bases were used: Symmlet 8 for \tt Heavisine \rm and \tt Doppler\rm, Daubechies 6 for \tt Bumps \rm and Haar for \tt Blocks\rm. The two columns of the design matrix were generated as independent $N(0,1)$ random variables.

Results of the simulation are presented in Table \ref{plm_table1}. It can be seen that the proposed \textit{GS-WaPaLiM} method gives better AMSE and AMSE$_{\bm{\beta}}$ results in most test scenarios. It is apparent that the modified version of the Backfitting algorithm (\textit{BFM}) provides better results than the original backfitting algorithm (\textit{BF}). Note that an additional uncertainty results from estimating the noise variance $\sigma^2$, which was assumed to be known in the simulations by \citet{Chang2004}. $LEGEND$ provides comparable results to the $BF$ algorithm, since both are using the same least squares formulation penalizing only the wavelet coefficients of the nonparametric part of the model. The solution algorithm and estimation of the noise is different in these methods. Note that boldface numbers indicate the smallest AMSE result for each test scenario.

\begin{table}[htbp]
\scriptsize
\caption[AMSE comparison of the \textit{GS-WaPaLiM} method to other methods for Example 1.]{\fontsize{11}{13}\selectfont AMSE comparison of the \textit{GS-WaPaLiM} method to other methods for Example 1.}

\label{plm_table1}
\centering
\resizebox{\linewidth}{!} {
\begin{tabular}[h]{|l|c|c||c|c|||l|c|c||c|c|}
	\hline
Signal & N & Method & AMSE & AMSE$_{\bm{\beta}}$ & Signal &  N & Method & AMSE & AMSE$_{\bm{\beta}}$ \\
	\hline \hline
Blocks     & 64   &GS-WaPaLiM &  \bf     0.6012  &  \bf     0.1179  &
Doppler    & 64   &GS-WaPaLiM &  \bf     1.0332  &          0.2009  \\
           &      &BF         &          8.3137  &          0.5647  &
           &      &BF         &          5.0350  &          0.3366  \\
           &      &BFM        &          1.0670  &          0.1606  &
           &      &BFM        &          1.1988  &          0.1607  \\
           &      &LEGEND     &          7.0360  &          0.4970  &
           &      &LEGEND     &          4.8372  &          0.2965  \\
           &      &DPPLM-GCV  &          0.8781  &          0.1415  &
           &      &DPPLM-GCV  &          1.0435  &  \bf     0.1535  \\ \hline
           & 128  &GS-WaPaLiM &  \bf     0.3933  &  \bf     0.0284  &
           & 128  &GS-WaPaLiM &  \bf     0.4865  &  \bf     0.0363  \\
           &      &BF         &          3.9360  &          0.1268  &
           &      &BF         &          2.5299  &          0.0758  \\
           &      &BFM        &          0.6040  &          0.0368  &
           &      &BFM        &          0.6676  &          0.0390  \\
           &      &LEGEND     &          3.6166  &          0.1166  &
           &      &LEGEND     &          2.8607  &          0.0811  \\
           &      &DPPLM-GCV  &          0.5955  &          0.0372  &
           &      &DPPLM-GCV  &          0.6540  &          0.0393  \\ \hline
           & 256  &GS-WaPaLiM &  \bf     0.2547  &  \bf     0.0107  &
           & 256  &GS-WaPaLiM &  \bf     0.3727  &  \bf     0.0126  \\
           &      &BF         &          2.1635  &          0.0320  &
           &      &BF         &          1.7494  &          0.0264  \\
           &      &BFM        &          0.4488  &          0.0138  &
           &      &BFM        &          0.4880  &          0.0145  \\
           &      &LEGEND     &          1.9638  &          0.0290  &
           &      &LEGEND     &          1.8516  &          0.0261  \\
           &      &DPPLM-GCV  &          0.4465  &          0.0140  &
           &      &DPPLM-GCV  &          0.4854  &          0.0146  \\ \hline
           & 512  &GS-WaPaLiM &  \bf     0.1776  &  \bf     0.0048  &
           & 512  &GS-WaPaLiM &  \bf     0.2293  &  \bf     0.0050  \\
           &      &BF         &          1.2914  &          0.0098  &
           &      &BF         &          0.9649  &          0.0085  \\
           &      &BFM        &          0.3247  &          0.0056  &
           &      &BFM        &          0.3129  &          0.0057  \\
           &      &LEGEND     &          1.2862  &          0.0096  &
           &      &LEGEND     &          1.0617  &          0.0089  \\
           &      &DPPLM-GCV  &          0.3252  &          0.0057  &
           &      &DPPLM-GCV  &          0.3122  &          0.0057  \\ \hline
Bumps      & 64   &GS-WaPaLiM &  \bf     0.7932  &  \bf     0.2136  &
Heavisine  & 64   &GS-WaPaLiM &  \bf     0.4265  &          0.0714  \\
           &      &BF         &          8.8940  &          0.6099  &
           &      &BF         &          1.4997  &          0.0959  \\
           &      &BFM        &          1.7003  &          0.2479  &
           &      &BFM        &          0.5890  &  \bf     0.0628  \\
           &      &LEGEND     &          9.2580  &          0.5556  &
           &      &LEGEND     &          1.5195  &          0.1013  \\
           &      &DPPLM-GCV  &          1.4129  &          0.2222  &
           &      &DPPLM-GCV  &          0.5833  &          0.0642  \\ \hline
           & 128  &GS-WaPaLiM &  \bf     0.7265  &          0.0783  &
           & 128  &GS-WaPaLiM &  \bf     0.2834  &          0.0227  \\
           &      &BF         &          6.3618  &          0.2014  &
           &      &BF         &          0.4817  &          0.0234  \\
           &      &BFM        &          1.0466  &          0.0728  &
           &      &BFM        &          0.3526  &  \bf     0.0218  \\
           &      &LEGEND     &          7.5358  &          0.2046  &
           &      &LEGEND     &          1.0038  &          0.0336  \\
           &      &DPPLM-GCV  &          0.9931  &  \bf     0.0716  &
           &      &DPPLM-GCV  &          0.3544  &          0.0226  \\ \hline
           & 256  &GS-WaPaLiM &  \bf     0.5522  &  \bf     0.0177  &
           & 256  &GS-WaPaLiM &  \bf     0.1972  &  \bf     0.0099  \\
           &      &BF         &          4.0588  &          0.0571  &
           &      &BF         &          0.3603  &          0.0112  \\
           &      &BFM        &          0.7845  &          0.0247  &
           &      &BFM        &          0.2623  &          0.0105  \\
           &      &LEGEND     &          3.9100  &          0.0533  &
           &      &LEGEND     &          0.6559  &          0.0140  \\
           &      &DPPLM-GCV  &          0.7688  &          0.0243  &
           &      &DPPLM-GCV  &          0.2608  &          0.0107  \\ \hline
           & 512  &GS-WaPaLiM &  \bf     0.4317  &  \bf     0.0065  &
           & 512  &GS-WaPaLiM &  \bf     0.1310  &  \bf     0.0045  \\
           &      &BF         &          2.9271  &          0.0196  &
           &      &BF         &          0.2576  &          0.0049  \\
           &      &BFM        &          0.6022  &          0.0091  &
           &      &BFM        &          0.1758  &          0.0047  \\
           &      &LEGEND     &          2.9142  &          0.0189  &
           &      &LEGEND     &          0.4219  &          0.0058  \\
           &      &DPPLM-GCV  &          0.5999  &          0.0090  &
           &      &DPPLM-GCV  &          0.1756  &          0.0047  \\ \hline
\end{tabular}
}
\end{table}
\normalsize
~\\
\noindent {\bf Example 2} \\
The second example is based on a simulation example from \citet{Ding2011}. The simulated data are generated from
\ba
Y_i = \bm{x}^T_i\bm{\beta}+\bm{f}(t_i)+\varepsilon_i,~~i=1,\ldots,n,
\ea
where $\varepsilon_i \sim N(0,1)$ and $\bm{\beta}=(1.5,2,2.5,3,0,\ldots,0)'$ with $p=20$. The parametric part of the model is sparse, where only the first 4 regression variables are significant. The nonparametric test functions are $\bm{f}(t)=c_j \bm{f}_j(t),~j=1,2$, where $\bm{f}_1(t)=$ \tt PiecePoly \rm given in \citet{Nason1996} and $\bm{f}_2(t)=$ \tt Bumps\rm. We chose $c_1=9$ and $c_2=3$ to have reasonable signal-to-noise ratios (SNR).  The test functions were simulated at $n=128,256$ and $512$ points, and Daubechies 8 wavelet base were used in both cases of the test functions. Rows of the design matrix $\bm{x}^T_1,\ldots,\bm{x}^T_n$ were independently generated from 20-dimensional multivariate normal distribution with zero mean vector, variance 1 and pairwise correlation coefficient between consecutive elements of the rows $\rho=0.4$.

Results of the simulation are presented in Table \ref{plm_table2}. Note that boldface numbers indicate the smallest AMSE results for each test scenario. It can be seen that the proposed \textit{GS-WaPaLiM} method gives better AMSE and AMSE$_{\bm{\beta}}$ results in all test scenarios. In this example the parametric part of the model is sparse, therefore, the double penalized wavelet estimator is superior to the wavelet backfitting and $LEGEND$ algorithms, especially in estimating $\beta_i$s. Since the true $\bm{\beta}$ is a sparse vector, penalized estimation of the coefficients provides superior results as opposed to the $BF$, $BFM$ and $LEGEND$ methods, which only penalize the wavelet coefficients corresponding to the nonparametric part in the estimation procedure. Similarly to Example 1, $LEGEND$ provides comparable results to the $BF$ algorithm. The proposed \textit{GS-WaPaLiM} method provides superior performance both in estimating the overall signal and the linear regression coefficients compared to the non-Bayesian methods considered.

\begin{table}[htbp]
\scriptsize
\caption[AMSE comparison of the \textit{GS-WaPaLiM} method to other methods for Example 2.]{\fontsize{11}{13}\selectfont AMSE comparison of the \textit{GS-WaPaLiM} method to other methods for Example 2.}
\label{plm_table2}
\centering
\resizebox{\linewidth}{!} {
\begin{tabular}[h]{|l|c|c||c|c|||l|c|c||c|c|}
	\hline
Signal & N & Method & AMSE & AMSE$_{\bm{\beta}}$ & Signal &  N & Method & AMSE & AMSE$_{\bm{\beta}}$ \\
\hline \hline
PiecePoly  & 128  &GS-WaPaLiM &  \bf     0.2752  &  \bf     0.0638  &
Bumps      & 128  &GS-WaPaLiM &  \bf     0.6706  &  \bf     0.1431  \\
           &      &BF         &          0.4467  &          0.4159  &
           &      &BF         &          2.5308  &          1.3895  \\
           &      &BFM        &          0.4052  &          0.4057  &
           &      &BFM        &          1.0980  &          1.0239  \\
           &      &LEGEND     &          0.4297  &          0.4111  &
           &      &LEGEND     &          2.8666  &          1.4833  \\
           &      &DPPLM-GCV  &          0.3642  &          0.1933  &
           &      &DPPLM-GCV  &          0.8640  &          0.6745  \\ \hline
           & 256  &GS-WaPaLiM &  \bf     0.1840  &  \bf     0.0287  &
           & 256  &GS-WaPaLiM &  \bf     0.4844  &  \bf     0.0458  \\
           &      &BF         &          0.3418  &          0.1788  &
           &      &BF         &          1.8775  &          0.4596  \\
           &      &BFM        &          0.2647  &          0.1689  &
           &      &BFM        &          0.6826  &          0.3133  \\
           &      &LEGEND     &          0.3045  &          0.1734  &
           &      &LEGEND     &          1.8776  &          0.4584  \\
           &      &DPPLM-GCV  &          0.2397  &          0.0763  &
           &      &DPPLM-GCV  &          0.6349  &          0.2103  \\ \hline
           & 512  &GS-WaPaLiM &  \bf     0.1113  &  \bf     0.0126  &
           & 512  &GS-WaPaLiM &  \bf     0.3910  &  \bf     0.0182  \\
           &      &BF         &          0.2417  &          0.0808  &
           &      &BF         &          1.5650  &          0.1896  \\
           &      &BFM        &          0.1678  &          0.0759  &
           &      &BFM        &          0.5266  &          0.1253  \\
           &      &LEGEND     &          0.2090  &          0.0784  &
           &      &LEGEND     &          1.4961  &          0.1842  \\
           &      &DPPLM-GCV  &          0.1516  &          0.0321   &
           &      &DPPLM-GCV  &          0.5119  &          0.0798  \\ \hline
\end{tabular}
}
\end{table}

\subsection{Variable selection}
A distinguishing feature of the proposed algorithm is that it can be used for variable selection. The method proposed by \citet{Ding2011} was developed for variable selection, but in the Bayesian framework, the method proposed by \citet{Qu2006} is not able to perform this important task.

The proposed methodology can simply mimic the machinery of SSVS (stochastic search variable selection) by \citet{George1993}. Recall, that latent variable $\gamma_i$ indicates whether predictor $i$ should be included in the model or not. We can select the best subset of linear predictors by using Gibbs sampling to identify models with higher posterior probability $f(\bm{\gamma}|d)$. In the Gibbs sampling procedure we generate the sequence $\bm{\gamma}^{(1)}, \bm{\gamma}^{(2)},\ldots,\bm{\gamma}^{(l)}$ which converges to the posterior distribution $f(\bm{\gamma}|\bm{d})$. Simple calculation of the empirical frequency of $\bm{\gamma}$ or different strategies mentioned in \citet{George1993} can be used to identify the best subsets of predictors.

To illustrate this, we show how variable selection works on Example 2 from the previous section, using \tt Bumps \rm for the nonparametric component and $n=128$. Remember that $p=20$, therefore, there are $2^{20}$ candidate models. Table \ref{plm_table3} shows 10 models with the highest estimated posterior probability based on 20,000 runs (5,000 was burn-in) of the Gibbs sampling algorithm. We can see that the method identifies the true model with distinctively highest posterior probability, even for $n=128$. In case $n=256$, the estimated posterior probability of the true model is 0.8128.

\begin{table}[htbp]
\small
\caption[Subset models with highest estimated posterior probabilities.]{\fontsize{11}{13}\selectfont Subset models with highest estimated posterior probabilities.}
\label{plm_table3}
\centering
\begin{tabular}[h]{|c|c|}
	\hline
Variables & Posterior probability \\
\hline \hline
$x_1,x_2,x_3,x_4$                & 0.2885  \\
$x_1,x_2,x_3,x_4,x_{19}$         & 0.1020  \\
$x_1,x_2,x_3,x_4,x_9$            & 0.0646  \\
$x_1,x_2,x_3,x_4,x_6$            & 0.0321  \\
$x_1,x_2,x_3,x_4,x_{16}$         & 0.0304  \\
$x_1,x_2,x_3,x_4,x_{15}$         & 0.0271  \\
$x_1,x_2,x_3,x_4,x_9,x_{15}$     & 0.0236  \\
$x_1,x_2,x_3,x_4,x_{20}$         & 0.0197  \\
$x_1,x_2,x_3,x_4,x_{16},x_{19}$  & 0.0167  \\
$x_1,x_2,x_3,x_4,x_{15},x_{19}$  & 0.0164  \\ \hline
\end{tabular}
\end{table}

\section{Conclusions}  \label{sec_conclusions_gswapalim}
In this paper we proposed a wavelet-based method for estimation and variable selection in partially linear models. Because wavelets provide efficient representation for wide ranges of functions, the inference was conducted in the wavelet domain. A fully Bayesian approach was taken, in which a mixture prior was specified on both the parametric and nonparametric components of the model, unifying modeling approaches from both the Bayesian linear models and the wavelet shrinkage literature. Estimation and variable selection was performed by a Gibbs sampling procedure. It was shown through simulated examples that the methodology provides superior performance compared to the penalized least squares approach, most common in the existing literature.

The developed algorithm is efficient; however, the computational time considerably increases when the number of covariates in the linear part of the model grows. Another limitation is the usual assumptions of wavelet regression, that is, we assumed equally spaced sampling points without replicates for the nonparametric component, and the number of observations was assumed to be a power of two. This can be a limitation for analyzing real-world data sets, however, wavelet transforms extending these assumptions can be found in the literature, see for example \citet{Kovac2000}.

\section{Appendix}
\label{sec:appendix}
First we provide derivation of results (\ref{plm_marg}) and (\ref{plm_post}). The joint distribution $f(d^{\star},\theta|\sigma^2)$ using prior $p_1(\theta|\tau)$ is
\ba
\displaystyle
f(d^{\star},\theta|\sigma^2,\tau) &=& f(d^{\star}|\theta,\sigma^2)p_1(\theta|\tau)=\frac{1}{\sqrt{2\pi\sigma^2}}e^{-\frac{(d^{\star}-\theta)^2}{2\sigma^2}}
\frac{\tau}{2}e^{-\tau \abs{\theta}}\\
\displaystyle
&=& \frac{\tau}{2\sqrt{2\pi\sigma^2}}e^{-\frac{1}{2\sigma^2}
\{\theta^2-2\theta(-\tiny{\sign}(\theta)\sigma^2\tau+d^{\star})+{d^{\star}}^2\}} \\
&=& \frac{\tau}{2\sqrt{2\pi\sigma^2}}e^{-\frac{1}{2\sigma^2}
\{ \theta-(d^{\star}-\tiny{\sign}(\theta)\sigma^2\tau)^2 \} }-
e^{-\frac{1}{2\sigma^2} \{ -(d^{\star}-\tiny{\sign}(\theta)\sigma^2\tau)^2+{d^{\star}}^2 \} } \\
&=& \frac{\tau e^{\frac{\sigma^2\tau^2}{2}}e^{-\tiny{\sign}(\theta)d^{\star}\tau}}{2\sqrt{2\pi\sigma^2}}
e^{-\frac{1}{2\sigma^2}
\{\theta-(d^{\star}-\tiny{\sign}(\theta)\sigma^2\tau)\}^2} \\
&=&
\begin{cases}
\displaystyle
\frac{\tau e^{\frac{\sigma^2\tau^2}{2}}e^{-d^{\star}\tau}}{2\sqrt{2\pi\sigma^2}}e^{-\frac{1}{2\sigma^2}
[\theta-(d^{\star}-\sigma^2\tau)]^2}, & \theta \geq 0 \\\\ \displaystyle
\frac{\tau e^{\frac{\sigma^2\tau^2}{2}}e^{d^{\star}\tau}}{2\sqrt{2\pi\sigma^2}}e^{-\frac{1}{2\sigma^2}
[\theta-(d^{\star}+\sigma^2\tau)]^2}, & \theta < 0
\end{cases}.
\ea
The marginal distribution becomes
\ba
\displaystyle
m(d^{\star}|\sigma^2,\tau)&=&\int_{-\infty}^{\infty}f(d^{\star},\theta|\sigma^2,\tau)d\theta \\
&=& \frac{\tau}{2} e^{\frac{\sigma^2\tau^2}{2}} \bigg\{
e^{d^{\star}\tau} \int_{-\infty}^{0} \frac{1}{\sqrt{2\pi\sigma^2}} e^{-\frac{1}{2\sigma^2}[\theta-(d^{\star}+\sigma^2\tau)]^2} d\theta + \\
&& e^{-d^{\star}\tau} \int_{0}^{\infty} \frac{1}{\sqrt{2\pi\sigma^2}} e^{-\frac{1}{2\sigma^2}[\theta-(d^{\star}-\sigma^2\tau)]^2} d\theta \bigg\} \\
&=& \frac{\tau}{2} e^{\frac{\sigma^2\tau^2}{2}} \bigg\{e^{d^{\star}\tau}\Phi\left(\frac{-d^{\star}-\sigma^2\tau}{\sigma}\right)+
e^{-d^{\star}\tau}\Phi\left(\frac{d^{\star}-\sigma^2\tau}{\sigma}\right)
\bigg\}.
\ea
Combining the two equations above, we get the posterior as
\ba
&& h(\theta|d^{\star},\sigma^2,\tau) = \frac{f(d^{\star},\theta|\sigma^2,\tau)}{m(d^{\star}|\sigma^2,\tau)}= \\
&&
\begin{cases}
\displaystyle
\frac{e^{-d^{\star}\tau}}{e^{-d^{\star}\tau}\Phi\left(\frac{d^{\star}}{\sigma}-\tau\sigma\right)+
e^{d^{\star}\tau}\Phi\left(-\frac{d^{\star}}{\sigma}-\tau\sigma\right)}
\frac{1}{\sqrt{2\pi\sigma^2}} e^{-\frac{1}{2\sigma^2}[\theta-(d^{\star}-\sigma^2\tau)]^2}, & \theta \geq 0 \\\\
\displaystyle
\frac{e^{d^{\star}\tau}}{e^{-d^{\star}\tau}\Phi\left(\frac{d^{\star}}{\sigma}-\tau\sigma\right)+
e^{d^{\star}\tau}\Phi\left(-\frac{d^{\star}}{\sigma}-\tau\sigma\right)}
\frac{1}{\sqrt{2\pi\sigma^2}} e^{-\frac{1}{2\sigma^2}[\theta-(d^{\star}+\sigma^2\tau)]^2}, & \theta < 0
\end{cases}. \\
\ea
These results were also derived by \citet{Pericchi1992} and used by \citet{Johnstone2005a}.

Now we derive the results used for the Gibbs sampling algorithm of model (\ref{plm_model2}). To derive the full conditional distribution for a parameter of interest we look at the joint distribution of all the parameters and collect the terms which contain the desired parameter. Let us denote $\boldsymbol{d}=\{d_{jk}:j=J_0,\ldots,\log_2(n)-1,~k=0,\ldots,2^j-1\}$,
$\boldsymbol{\beta}=\{\beta_i:i=1,\ldots,p\}$, $\boldsymbol{\theta}=\{\theta_{jk}:j=J_0,\ldots,\log_2(n)-1,~k=0,\ldots,2^j-1\}$,
$\boldsymbol{\gamma}=\{\gamma_i:i=1,\ldots,p\}$, $\boldsymbol{z}=\{z_{jk}:j=J_0,\ldots,\log_2(n)-1,~k=0,\ldots,2^j-1\}$,
$\boldsymbol{\eps}=\{\eps_j:j=J_0,\ldots,\log_2(n)-1\}$ and
$\boldsymbol{v}=\{v_i:i=1,\ldots,p\}$. The joint distribution of the data and parameters for model in (\ref{plm_model2}) becomes
\ba
&& f(\boldsymbol{d},\boldsymbol{\beta},\boldsymbol{\theta},\boldsymbol{\gamma},\boldsymbol{z},q,\boldsymbol{\eps},
\boldsymbol{v},\sigma^2,\tau_{\theta},\eta^2)=
\left[\prod_{j,k} \frac{1}{\sqrt{2\pi\sigma^2}}e^{-\frac{1}{2\sigma^2}(d_{jk}-(\bm{U}_{\bm{\gamma}}\bm{\beta}_{\bm{\gamma}})_{jk}-
\theta_{jk})^2} \right] \cdot \\
&& \frac{1}{\Gamma(a_1){b_1}^{a_1}}(\sigma^2)^{-a_1-1}e^{-\frac{1}{\sigma^2}\frac{1}{b_1}}
\left[\prod_i \left\{ (1-\gamma_i)\delta_0+
\gamma_i\frac{1}{\sqrt{2\pi v_i\eta^2}}e^{-\frac{1}{2v_i\eta^2}\beta^2_i} \right\} \right]
\cdot \\
&& \left[\prod_{j,k} \left\{ (1-z_{jk})\delta_0+z_{jk}\frac{\tau_{\theta}}{2}
e^{-\tau_{\theta}\abs{\theta_{jk}}} \right\} \right]
\left[ \prod_i q^{\gamma_i}(1-q)^{(1-\gamma_i)} \right] \cdot \\
&& \left[\prod_{j,k} \eps_j^{z_{jk}}(1-\eps_j)^{(1-z_{jk})} \right]
\mbox{\bf{1}}\{0\leq\ q \leq1\} \left[\prod_j \mbox{\bf{1}}\{0\leq\eps_j\leq1\} \right]
\left[\prod_i e^{-v_i} \right] \cdot \\
&& \frac{1}{\Gamma(a_2){b_2}^{a_2}}(\eta^2)^{-a_2-1}e^{-\frac{1}{\eta^2}\frac{1}{b_2}}
\frac{1}{\Gamma(a_3){b_3}^{a_3}}\tau_{\theta}^{a_3-1}e^{-\tau_{\theta}/b_3}. \\
\ea
The full conditional distribution of parameters $\beta_i$ and $\gamma_i$ simply follows from \citet{Yuan2004} with using $\bm{Z} = \bm{d}-\bm{U}_{\bm{\gamma}^{[-i]},\gamma_i=0}\bm{\beta}_{\bm{\gamma}^{[-i]},\gamma_i=0}-\bm{\theta}$.

\noindent The full conditional distribution of $v_i$ is
\ba
p(v_i|\beta_i,\gamma_i,\eta^2) &\propto& \left\{ (1-\gamma_i)\delta_0+
\gamma_i\frac{1}{\sqrt{2\pi v_i\eta^2}}e^{-\frac{1}{2v_i\eta^2}\beta^2_i}
\right\} e^{-v_i} \\
&=&
\begin{cases}
{\cal E}xp(1), & \mbox{if} \quad \gamma_i=0 \\
{\cal GIG} \left(2,\beta^2_i/\eta^2,1/2 \right), & \mbox{if} \quad \gamma_i=1
\end{cases},
\ea
where ${\cal GIG}(a,b,p)$ denotes the generalized inverse Gaussian distribution \citep[p.284]{Johnson1994} with probability density function
\ba
f(x|a,b,p)=\frac{(a/b)^{p/2}}{2K_p(\sqrt{ab})}x^{p-1}e^{-(ax+b/x)/2}, \quad x>0;a,b>0,
\ea
where $K_p$ denotes the modified Bessel function of the third kind.
\\
\noindent The full conditional distribution of $\eta^2$ is
\ba
\displaystyle
p(\eta^2|\boldsymbol{\beta},\boldsymbol{\gamma},\boldsymbol{v}) &\propto&
\left[\prod_i \left\{ (1-\gamma_i)\delta_0+
\gamma_i\frac{1}{\sqrt{2\pi v_i\eta^2}}e^{-\frac{1}{2v_i\eta^2}\beta^2_i} \right\} \right] \cdot \\
&& \frac{1}{\Gamma(a_2){b_2}^{a_2}}(\eta^2)^{-a_2-1}e^{-\frac{1}{\eta^2}\frac{1}{b_2}} \\
&\propto& (\eta^2)^{-a_2-1/2\sum_i \gamma_i-1} \exp\left\{-\frac{1}{\eta^2} \left( 1/b_2 +
1/2 \sum_i{\left( \gamma_i \beta^2_i / v_i \right)}
\right) \right\} \\
&=& {\cal IG} \left( a_2+1/2 \sum_i \gamma_i,
\left[1/b_2+1/2 \sum_i{\left( \gamma_i \beta^2_i / v_i \right)} \right]^{-1} \right). \\
\ea

\noindent The full conditional distribution of $q$ can be derived as
\ba
\displaystyle
p(q|\boldsymbol{\gamma}) &=& \left[ \prod_i q^{\gamma_i}(1-q)^{(1-\gamma_i)} \right] \mbox{\bf{1}}\{0\leq q \leq1\} \\
&\propto& q^{\sum_i \gamma_i}(1-q)^{p-\sum_i \gamma_i} \mbox{\bf{1}}\{0\leq q \leq1\} \\
&=& {\cal B}e \left(1+\sum_i \gamma_i, 1+\sum_i \left(1-\gamma_i\right) \right). \\
\ea

\noindent The full conditional distribution of $\sigma^2$ is
\ba
\displaystyle
&& p(\sigma^2|\boldsymbol{\beta},\boldsymbol{\theta},\boldsymbol{\gamma},\boldsymbol{d}) \propto (\sigma^2)^{-n/2}e^{-\frac{1}{2\sigma^2}\sum_{j,k}(d_{jk}-(\bm{U}_{\bm{\gamma}}\bm{\beta}_{\bm{\gamma}})_{jk}-
\theta_{jk})^2} (\sigma^2)^{-a_1-1}e^{-\frac{1}{\sigma^2}\frac{1}{b_1}} =\\
&& (\sigma^2)^{-a_1-n/2-1}\exp\left\{-\frac{1}{\sigma^2}\left(1/b_1+1/2\sum_{j,k}\left(d_{jk}-
(\bm{U}_{\bm{\gamma}}\bm{\beta}_{\bm{\gamma}})_{jk}-\theta_{jk}\right)^2 \right)\right\}= \\
&& {\cal IG} \left( a_1+n/2,
\left[1/b_1+1/2\sum_{j,k}\left(d_{jk}-(\bm{U}_{\bm{\gamma}}\bm{\beta}_{\bm{\gamma}})_{jk}-\theta_{jk}\right)^2 \right]^{-1} \right). \\
\ea
In the following, we denote $d^{\star}_{jk}=d_{jk}-(\bm{U}_{\bm{\gamma}}\bm{\beta}_{\bm{\gamma}})_{jk}$. The conditional distribution of $z_{jk}$ remains Bernoulli with posterior probability derived by
\ba
\displaystyle
&& P(z_{jk}=1|d^{\star}_{jk},\sigma^2,\tau,\eps_j) = \frac{P(z_{jk}=1|\eps_j)f(d^{\star}_{jk}|\sigma^2,\tau,z_{jk}=1)}
{\sum_{i\in\{0,1\}} P(z_{jk}=i|\eps_j)f(d^{\star}_{jk}|\sigma^2,\tau,z_{jk}=i)}= \\
&& \frac{P(z_{jk}=1|\eps_j) \int_{-\infty}^{\infty} f(d^{\star}_{jk}|\theta_{jk},\sigma^2)p(\theta_{jk}|\tau,z_{jk}=1) d\theta_{jk}}
{\sum_{i\in\{0,1\}} P(z_{jk}=i|\eps_j) \int_{-\infty}^{\infty} f(d^{\star}_{jk}|\theta_{jk},\sigma^2)p(\theta_{jk}|\tau,z_{jk}=i) d\theta_{jk}}= \\
&& \frac{\eps_j m\left(d^{\star}_{jk}|\sigma^2,\tau \right)}
    {\left(1-\eps_j\right)f\left(d^{\star}_{jk}|0,\sigma^2\right)+
    \eps_j m\left(d^{\star}_{jk}|\sigma^2,\tau \right)}. \\
\ea
Here $p(\theta_{jk}|\tau,z_{jk}=i),~i\in\{0,1\}$ denote the two parts of the mixture prior in model (\ref{plm_model2}), depending on the value of latent variable $z_{jk}$. Similar result for the full conditional of $\gamma_i$ was used by \citet{Yuan2005}.
\\\\
\noindent The full conditional distribution of $\eps_j$ is
\ba
\displaystyle
p(\eps_j|\boldsymbol{z}) &\propto& \left[\prod_{k} \eps_j^{z_{jk}}(1-\eps_j)^{(1-z_{jk})} \right] \mbox{\bf{1}}\{0\leq\eps_j\leq1\} \\
&=& \eps_j^{\sum_k z_{jk}}(1-\eps_j)^{\sum_k(1-z_{jk})} \\
&=& {\cal B}e \left(1+\sum_k z_{jk}, 1+\sum_k \left(1-z_{jk}\right) \right). \\
\ea
Similarly, the full conditional distribution of $\theta_{jk}$ is
\ba
\displaystyle
p(\theta_{jk}|d_{jk},\boldsymbol{\beta},\boldsymbol{\gamma},z_{jk},\sigma^2,\tau_{\theta}) &\propto& \exp\left\{-\frac{1}{2\sigma^2}(d_{jk}-(\bm{U}_{\bm{\gamma}}\bm{\beta}_{\bm{\gamma}})_{jk}-\theta_{jk})^2\right\} \cdot \\
&& \left\{ (1-z_{jk})\delta_0+z_{jk}\frac{\tau_{\theta}}{2}e^{-\tau_{\theta}\abs{\theta_{jk}}} \right\} \\
&=&
\begin{cases}
\delta_0(\theta_{jk}), & \mbox{if} \quad z_{jk}=0 \\
h(\theta_{jk}|d^{\star}_{jk},\sigma^2,\tau_{\theta}), & \mbox{if} \quad z_{jk}=1
\end{cases},
\ea
where the distribution $h(\theta_{jk}|d^{\star}_{jk},\sigma^2,\tau_{\theta})$ comes from the result in (\ref{plm_post}) and was derived above.\\
\\ Finally, the full conditional distribution of $\tau_{\theta}$ is
\ba
\displaystyle
p(\tau_{\theta}|\boldsymbol{\theta},\boldsymbol{z}) &\propto&
\left[ \prod_{j,k} \left\{ (1-z_{jk})\delta_0+z_{jk}\frac{\tau_{\theta}}{2}\exp\left(-\tau_{\theta}\abs{\theta_{jk}}\right) \right\} \right] \cdot \\
&& \frac{1}{\Gamma(a_3){b_3}^{a_3}}\tau_{\theta}^{a_3-1}\exp\left(-\tau/b_3 \right) \\
&\propto& \tau_{\theta}^{a_3+\sum_{j,k}z_{jk}-1}\exp\left\{-\tau_{\theta} \left(\sum_{j,k} \left(z_{jk}\abs{\theta_{jk}}\right)+1/b_3 \right) \right\} \\
&=& {\cal G}a \left(a_3+\sum_{j,k}z_{jk},\left[1/b_3+\sum_{j,k}\left(z_{jk}\abs{\theta_{jk}}\right)
\right]^{-1} \right). \\
\ea

\bibliographystyle{Chicago}

\bibliography{thesis-references}
\end{document}